\documentclass[%
 aip,
 jmp,%
 amsmath,amssymb,
preprint,%
]{revtex4-1}

\usepackage{graphicx}
\usepackage{dcolumn}
\usepackage{bm}

\begin{document}

\preprint{AIP/123-QED}

\title{New Multi-Scale Simulation Framework for Next-Generation Electronic Design Automation with Application to the Junctionless Transistor}

\author{J. Peng}
\affiliation{%
Department of Chemistry, The University of Hong Kong,
Hong Kong, China
}%

\author{Q. Chen}
\affiliation{Department of Electrical and Electronic Engineering, The University of Hong Kong,
Hong Kong, China}

\author{N. Wong}
\affiliation{Department of Electrical and Electronic Engineering, The University of Hong Kong,
Hong Kong, China}

\author{L. Y. Meng}
\affiliation{%
Department of Chemistry, The University of Hong Kong,
Hong Kong, China
}%

\author{C. Y. Yam}
\affiliation{%
Department of Chemistry, The University of Hong Kong,
Hong Kong, China
}%

\author{G. H. Chen}
\email{ghc@yangtze.hku.hk}
 \homepage{http://yangtze.hku.hk/home/}
\affiliation{%
Department of Chemistry, The University of Hong Kong,
Hong Kong, China
}%

\date{\today}

\begin{abstract}
In this paper we present a new multi-scale simulation scheme for next-generation electronic design automation for nano-electronics. The scheme features a combination of the first-principles quantum mechanical  calculation, semi-classical semiconductor device simulation, compact model generation and circuit simulation. To demonstrate the feasibility of the proposed scheme, we apply our newly developed quantum mechanics/electromagnetics method to simulate the junctionless transistors. The simulation results are consistent with the experimental measurements and provide new insights on the depletion effect of the hetero-doped gate on the drain current. Based on the calculated I-V curves, a compact model is then constructed for the junctionless transistors. The validity of the compact model is further verified by the transient circuit simulation of an inverter.
\end{abstract}

\pacs{85.35.-p, 73.63.-b}
\keywords{Junctionless, QM/EM, Multi-scale}
\maketitle

\section{Introduction}

The continuous miniaturization of semiconductor devices has resulted in the $22$nm transisters and is expected to approach $10$nm and beyond before 2026~\cite{ITRS:11}. Atomic features and quantum effects have become more pronounced than ever before, and some are even exploited as functioning mechanism of a range of new devices, such as transistors based on quantum dots~\cite{Devoret:2000fk} and wires~\cite{Cui:03} . The major challenges of the existing electronic design automation (EDA) tools are in how to incorporate the quantum phenomena into the classical semiconductor device models based on the continuum assumption. Full atomistic quantum mechanical (QM) calculation, while providing us with reliable characterizations of nano-scale systems, confronts severe applicability limitation due to the prohibitive computational cost. One natural solution is the multi-scale method, where only a small portion of the device is characterized with QM description and the remaining (non-critical) parts of the device plus the surrounding structures are treated classically. The rationale behind lies in that the QM effects that have substantial influence on device properties are confined to a small region, outside which (e.g., in the bulk material) classical models suffice to provide reasonable characterizations.

One element of this programme, namely a multi-scale quantum mechanics/electromagnetics (QM$/$EM) simulation framework that combines a first-principles QM simulator and a semi-classical EM solver, has already been delivered by the LODESTAR code~\cite{yam2011,meng2012}, which achieves a high performance-cost-ratio that cannot be achieved otherwise. The QM/EM method has been applied to simulate the carbon nanotubes embedded in silicon substrates, and the good agreements with full QM simulation have been obtained.

Physics-based simulation of individual devices, such as the QM and QM/EM simulations mentioned above, comprises only one part in the modern EDA flow. One level upward the technology hierarchy, compact modeling of device is another necessity to enable computer simulation involving millions of devices in circuit simulation programme, such as SPICE~\cite{SPICE}. State-of-the-art compact models of transistors, such as the BSIM family, are based on a process called  parameter extraction, to determine a set of model parameters by fitting the measured I-V curves. The experimental data required to generate compact models are usually costly to obtain, in terms of both time and expense, particularly for emerging devices and structures for which experiments may not be available or mature enough to provide reliable data.

The objective of this paper is to report a multi-scale flow for next-generation EDA tools that spans from the first-principles QM simulation to the generation of the compact models for circuit simulations. To the knowledge of the authors, it is the first time that such a  flow is reported in the literature. To demonstrate the feasibility of the entire flow, we apply the QM/EM method to the simulation of a new type of semiconductor device, the junctionless transistor, and generate a compact model based on the QM/EM results, which is then put into a circuit simulator to simulate a simple integrate circuit made of the junctionless transistors.

\section{QM/EM Simulation}

\subsection{Basic Framework}
The QM/EM method starts with partitioning the system of interest into QM regions which are described by quantum mechanics, and EM regions where classical models will be used. Typical domains that should be treated at QM level in a nanotransistor include the channel and parts of the source and drain for conduction currents. We did an external QM calculation with the gate included to demostrate the quantum depletion effect caused by the hetero-doped gate. The EM regions consist the remaining parts of the system.

The QM part of the system is solved by using our first-principles simulator {\emph{LODESTAR}}~\cite{lodestar}, with the density-functional tight-binding (DFTB) method combined with the non-equilibrium Green's function (NEGF) method~\cite{Pecchia:2008fk}. The EM solver combines the Maxwell's equations and the drift-diffusion (DD) model, as described in references ~\cite{Wim:01,Wim:02}. More advanced technology computer-aided design (TCAD) models for semiconductor devices can be readily used to replace the DD model for higher modeling requirements.

The two solvers are coupled via boundary conditions at the interfaces of the QM and EM regions. The EM solver is first applied in the whole domain to obtain an initial potential distribution. The potentials at the QM/EM interface then act as the boundary conditions for the QM calculation. In return, the current density through the QM/EM interface calculated by the QM solver is served as a part of the boundary conditions for the EM solver. The process is iterated until the current and potential at the interface converge. For more details we refer the readers to our previous papers~\cite{yam2011,meng2012}.

\subsection{Application on Junctionless Transistor}
In this section, we apply the QM/EM method to simulate a new type of semiconductor device, junctionless transistor. The junctionless transistor revives in responding to the increasing difficulty in fabricating P-N junctions in devices smaller than $10$nm, which requires an abrupt doping variation (e.g., from P-type to N-type) within a few nanometers and results in high electrical interference between source and drain. Basing on the ideas of Lilienfeld in the 1920s~\cite{lilienfeld1925} and more recent calculations\cite{soree2008},  Colinge {\emph et al.} presented the first experimental realization of the junctionless transistor in 2010~\cite{colinge2010}. The channel of the device is made of an uniformly-doped silicon nanowire, and a hetero-doped gate, viz., the gate doped with different type of dopant from the channel, is used to deplete the carriers inside the channel and keeps the transistor ``OFF'' in the absence of an applied gate voltage. Positive or negative gate voltage is required to turn ``ON'' the device. Detailed measurements~\cite{lee2010a,lee2010b,Raskin,Akhavan2011,Yan2011}, as well as atomistic scale simulations \cite{Ansari2010, Ansari2011} were followed up by the same group. The success of the junctionless transistor has attracted great attention of both experimentalists and theorists~\cite{Martinez2011,Gnani2011,Sels2011,Pham2011,park:2012qr} in the world.

\begin{figure}[ht]
\includegraphics[width=\columnwidth]{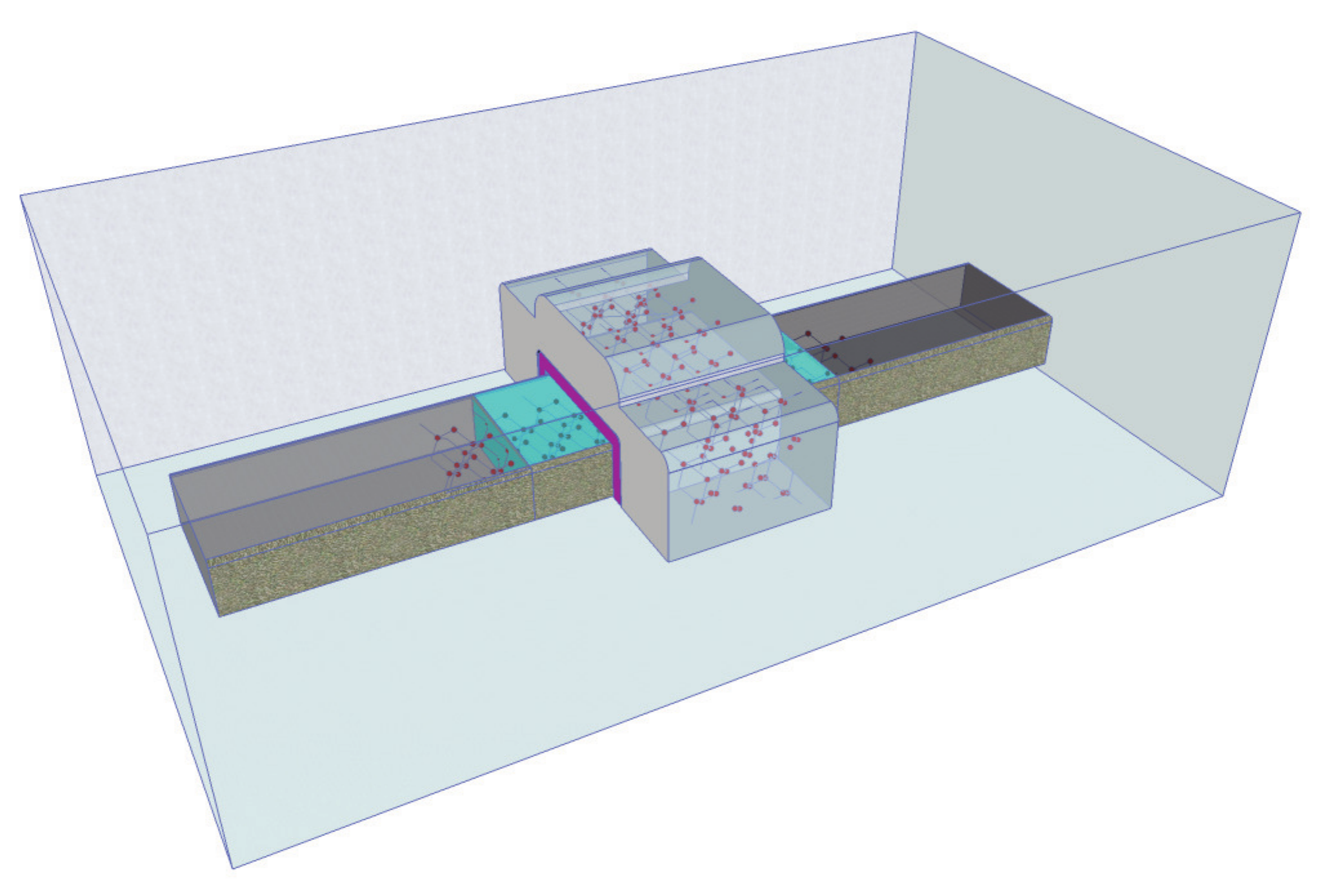}
\caption{The schematic representation of the QM/EM structure. QM/EM region is devided differently in the ground state and steady state calculation. A 6 nm long silicon nanowire with the cross section of  $2.0\times1.5$ $nm^2$ doped by 6 Ga/As atoms and the $\Pi$ shaped gate above it are included in the ground state QM calculation; When the QM/EM method is applied to the steady state calculation,  the QM region is further cutted down to a 3 nm nanowire ( shown as blue region). EM part: The remaining part of the $8\times6\times6$ $nm^3$ EM box.\label{fig:qmemstruct}}
\end{figure}

The structure of the junctionless transistor is shown in Fig.~\ref{fig:qmemstruct}. The QM/EM region is divided differently in the ground state and steady state calculation. The QM region for ground state calculation includes a $6$nm-long silicon nanowire doped by 6 Ga/As atoms along the (110) direction, which includes the source, channel and drain, and a ``$\Pi$'' shaped hetero/homo-doped gate (the gate doped with the same/different type of dopants). From this calculation we obtain the free particles distribution in the absence of the applied gate voltage and the bias voltage, which gives us the figures of the depletion effect; When the QM/EM method is applied to the more time consuming steady state calculation, the QM region is further cutted down to a 3 nm long silicon nanowire, as shown in Fig.~\ref{fig:qmemstruct}, the blue region. The EM region consists the remaining parts of the structure in the EM box with the size of $8\times6\times6$ $nm^3$, which is very difficult to be handled as a whole by the exists atomistic QM simulators.

The depletion effect of a hetero-doped gate was characterized in the previous reports through an artificial work function difference between the transistor and the electrons to shift the ``OFF'' state to zero gate voltage both classically and quantum mechanically~\cite{lee2010b,Gnani2011}, where the value of the ``shifting'' is simply given as the difference of the Fermi levels of the N- and P-doped bulk silicon. In reality, the depletion effect is more complicated than the simple shifting of the Fermi level in the channel area. With explicit QM account of the gate, we are able to calculate the effect of the intrinsic work function difference more accurately.

\begin{figure}[ht]
\begin{center}
\includegraphics[width=\columnwidth]{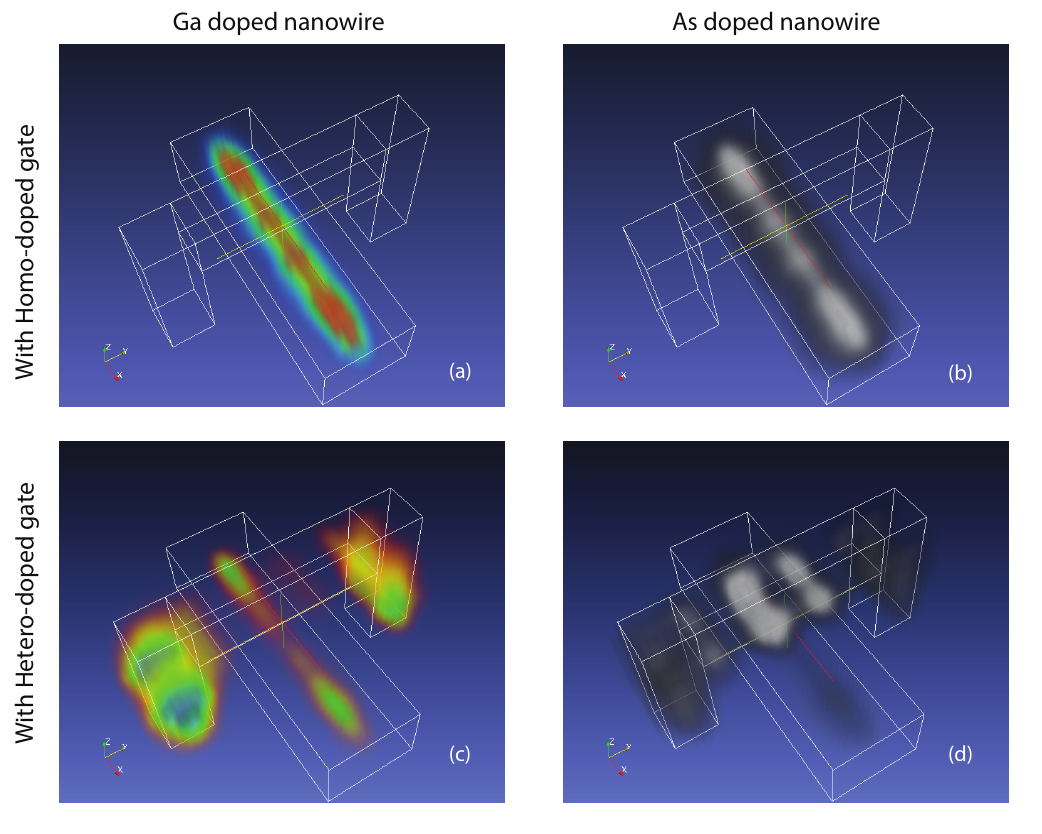}
\caption{Free electrons (6 in total) distribution in the N-channel nanowire with homo-doped gate (a) and with hetero-doped gate (c) ; free holes (6 in total) distribution in the P-channel nanowire with homo-doped gate (b) and with hetero-doped gate (d). In both (c) and (d), the free particles have been depleted by the hetero-doped gate. \label{fig:charge}}
\end{center}
\end{figure}

In order to demonstrate the impact of the gate doping, we simulate four structures with the same geometry but different doping combinations, and plot in Fig.~\ref{fig:charge} the $6$ electrons which occupy the highest occupied states in the N-channel devices (a,c) and the $6$ holes which occupy the lowest unoccupied states (b,d). A temperature of $6$K is introduced to help the convergence, resulting in a certain occupation above the Fermi level. The distribution of the free electrons in the N-channel device is obtained by $\rho_e(r)=\sum_{N/2-2}^NO_i\phi^*_i(r)S\phi_i(r)$, and can be seen in Fig. ~\ref{fig:charge} (a) and (c). And the distribution of the free holes in the P-channel device is obtained by $\rho_h=\sum_{1}^{N/2+2}(2.0-O_i)\phi^*_iS\phi_i$ which has been seen in Fig. ~\ref{fig:charge} (b) and (d). Here $N$ is the number of electrons, $S$ the overlap matrix, $\phi_i$ the eigenfunction and $O_i$ the occupation.

Fig.~\ref{fig:charge} shows that, with homo-doped gate, no carrier depletion occurs, and the device is normally ``ON'' without gate bias. With hetero-doped gate, on the other hand, a significant portion of the carriers in the silicon nanowire is depleted to the gate, which results in the lack of carriers inside the channel and a great reduction in the conductivity of the transistor in the absence of the applied gate voltage. Positive or negative gate voltage is required to ``push'' the free carriers back to the channel and turn on the device. By controlling the doping type of the gate, one can construct either a depletion-mode transistor, which is ``ON'' at zero bias, or an enhancement-mode transistor, which is ``OFF'' at zero bias.

\begin{figure}
\includegraphics[width=\columnwidth]{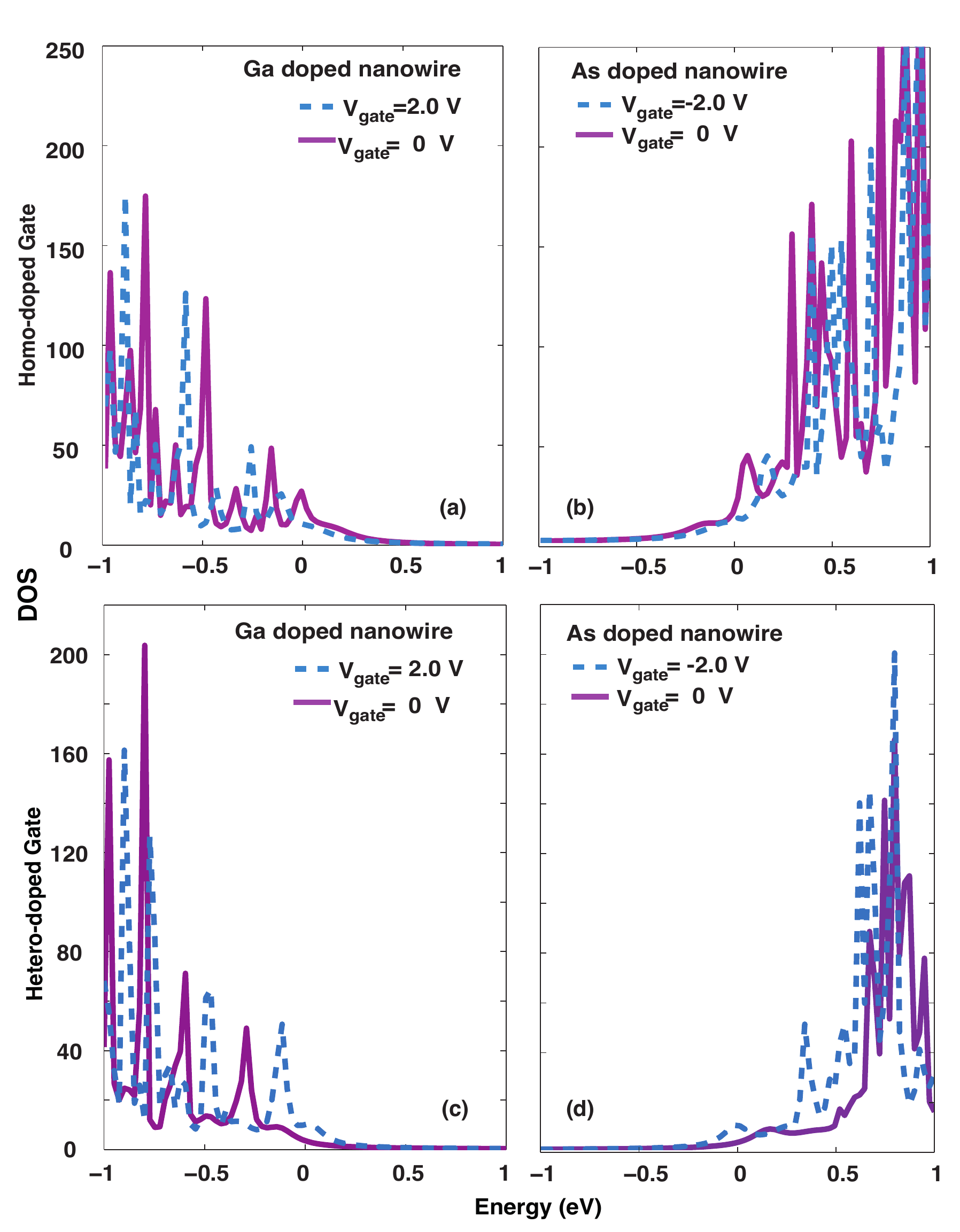}
\caption{Density of States (a),(b),(c),(d) correspond to structure (a),(b),(c)and(d) in Fig. \ref{fig:charge}.  The purple solid lines represent the DOS without applied gate voltage, which are rich in figures (a) and (b) while, because of the hetero-doped  gates, poor in figures (c) and (d) around the Fermi level. Positive or negative gate voltage is applied to shift the DOS to the blue dotted lines hence the ON/OFF states of the devices are changed. \label{fig:dos}}
\end{figure}

We also simulate the density of state (DOS) of the channel region. DOS in the absence of gate voltage and in a certain gate voltage has been obtained for the four different doped structures, as illustrated in Fig.~\ref{fig:dos} (a), (b), (c) and (d) corresponding to the structures (a), (b), (c) and (d) in Fig.~\ref{fig:charge}. The atomic structures of the QM regions of (a) and (c) are identical while the calculated Fermi levels are different, resulted from the depletion effect of the hetero-doped gate which attracts the carriers to the gate (see Fig.~\ref{fig:charge} (c) and (d)). The difference near the Fermi level is clearly shown in  Fig.~\ref{fig:dos} (a) and (c) when comparing the solid purple lines that represent the DOS in the zero gate bias . Similar analysis can be applied in Fig.~\ref{fig:dos} (b) and (d). The DOS figures show the electronic structures insights of the ``ON/OFF'' character of the transistors.

\begin{figure}[ht]
\includegraphics[width=\columnwidth]{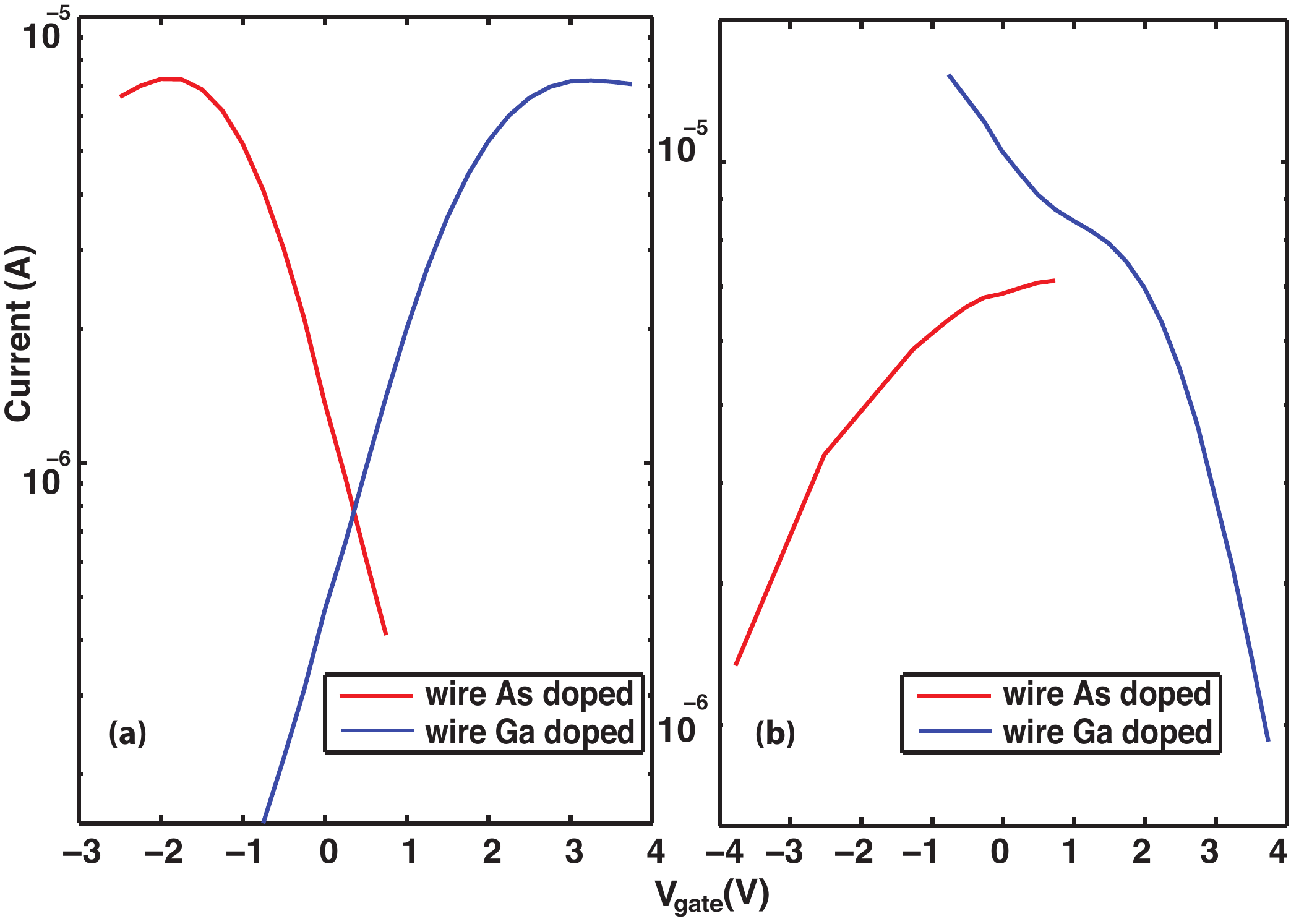}
\caption{ $I-V_{gate}$ curves for hetero-doped structures (a) and homo-doped structures (b).  \label{fig:current}}
\end{figure}

To generate the compact model for the junctionless transistor, we simulate the $I-V_{gate}$ curves for the hetero-doped and the homo-doped structures under the source-drain bias of 0.2V, and display them in Fig. ~\ref{fig:current}. Since multiple simulations are needed to sample the gate voltages, and only the transport dynamics in the channel are of interest, we model the gate region with the EM simulator to reduce the burden of QM solver and speed up the computation. In the hetero-doped structure, the conducting current in channel is very low in the absence of a gate bias due to the lack of free carriers. When positive/negative gate voltage is applied, the carriers ``absorbed'' by the gate are driven back to the channel and switch the device from ``OFF'' to ``ON''. On the contrary, the device with homo-doped gate is normally ``ON'' at zero gate voltage and is turned ``OFF'' when a gate voltage is applied to squeeze the carriers out of the channel.

Although the $I-V_{gate}$ curves are qualititatively consistent with the experimental results~\cite{colinge2010}, the magnitude of the ``OFF'' current is several orders larger than the curves in the measurements. The reason lies in the limited size of the QM region that can be handled by our first-principles simulator. Two dopants in a $3\times2\times1.5$ $nm^3$ silicon wire leads to an equivalent doping concentration of $2.6\times 10^{20}$cm$^{-3}$, which is about one order higher than the ordinary values $2\sim 5\times 10^{19}$cm$^{-3}$ for junctionless transistors. Such heavy doping density renders the silicon wire behave more like a conductor, and therefore, leads to relatively large current even at the ``OFF'' state. Meanwhile, the ultrashort gate length $\sim 1$ nm may cause the quantum barrier not large enough to prevent the electrons from tunneling through the channel, which can also result in the large ``OFF'' current.

\section{Compact Modeling}
Parameter extraction is the essential part of compact modeling for semiconductor devices. The ultimate objective of our method is to replace the experimental data required in the parameter extraction process by the QM/EM simulation results. A direct connection between first-principles simulation and SPICE-compatible compact models helps to minimize costly experiment measurements and facilitate the modeling of emerging semiconductor devices from first-principles.

To demonstrate the multi-scale QM/EM-to-SPICE modeling process, we generate a primitive compact model for junctionless transistor out of the I-V curves shown in Fig.~\ref{fig:current} (a). Our compact model formulation is based upon a recently developed model~\cite{Duarte:11} for the junctionless transistors. The operations of junctionless transistor are partitioned as linear, saturation and subthreshold regions according to the gate bias. The drain-to-source currents $I_{ds}$ are modeled separately in the three regions. Several modifications are made to the original model~\cite{Duarte:11} to render it suitable for our problem.

In the linear region, where the gate-to-source voltage $V_{gate}$ is larger than the threshold voltage $V_{th}$ and the drain-to-source voltage is small ($V_{gate} <V_{gate}-V_{th}$), the current through the device $I_{ds}$ is linearly proportional to $V_{gate}$
\begin{equation}\label{eq:curr_lin}
    {I_{ds}} \approx 2{\mu}\frac{{{\varepsilon _{ox}}}}
{{\beta {t_{ox}}}}\frac{W}
{L}\left( {{V_{gate}} - {V_{th}} +\Delta V_{th} - \frac{{{V_{ds}}}}
{2}} \right){V_{ds}}
\end{equation}
where $\beta = 1+\varepsilon_{ox}t_{si}/(4\varepsilon_{si}t_{ox})$ with $\varepsilon_{si}$ and $\varepsilon_{ox}$ being the permittivity of silicon and silicon oxide, and $t_{si}$ and $t_{ox}$ being the thickness of the silicon channel and the thickness of the silicon oxide layer between the gate and the channel, respectively. $W$ and $L$ are respectively the width and length of the device, and $N_{si}$ is the doping concentration of the channel. The threshold voltage $V_{th}$ is determined by
\begin{equation}\label{eq:Vth}
    {V_{th}} = {V_{fb}} - \frac{{q{N_{si}}{t_{si}}{t_{ox}}}}
{{2{\varepsilon _{ox}}}} - \frac{{q{N_{si}}t_{si}^2}}
{{8{\varepsilon _{si}}}}
\end{equation}
where $V_{fb}$ is the flat band voltage of silicon. Since the structure in our simulation is ultra short and narrow, we introduce $\Delta V_{th}$ as a shift of the threshold voltage to take into account the short channel effect (SCE) and drain induced barrier lowering (DIBL)~\cite{Tang:09}. The mobility $\mu$ is calculated by
\begin{equation}\label{eq:mu_lin}
    \mu=s_{lin}\frac{L}{W}\frac{1}{V_{ds}}\frac{1}{C_i}
\end{equation}
in which $s_{lin}$ is the slope of I-V curve measured in the linear region and $C_i = 2\varepsilon_{ox}/(t_{ox}\beta)$ is the gate insulator capacitance per unit area.

When $V_{ds} \gg V_{gate}-V_{th}>0 $, the I-V curve enters the saturation region, where the drain current is expressed by
\begin{equation}\label{eq:curr_sat}
    {I_{ds}} = \mu \frac{{{\varepsilon _{ox}}}}
{{\beta {t_{ox}}}}\frac{W}
{L}\left[ {{{\left( {{V_{gate}} - {V_{th}}+\Delta V_{th}} \right)}^2} - \frac{{{v_T}\beta {t_{ox}}}}
{{{\varepsilon _{ox}}}}\sqrt {2q{N_{si}}\pi {v_T}{\varepsilon _{si}}} {e^{\left( {{V_{gate}} - {V_{th}}+\Delta V_{th} - {V_{ds}}} \right)/{v_T}}}} \right]
\end{equation}
where $v_{T}$ is the thermal voltage $kT/q$.

The subthreshold region (or the ``OFF'' region) occurs when $V_{gate}<V_{th}$. In this region, $I_{ds}$ is described by
\begin{equation}\label{eq:curr_sub}
    {I_{ds}} = {\mu _{sub}}{v_T}\frac{W}
{L}\sqrt {2q{N_{si}}\pi {v_T}{\varepsilon _{si}}} {e^{\left( {{V_{gate}} - {V_{th}}}+\Delta V_{th} \right)/\left( {{v_T}{n_{slope}}} \right)}}\left( {1 - {e^{ - {V_{ds}}/{v_T}}}} \right)
\end{equation}
Here a different mobility is applied in the sub-threshold region $\mu_{sub}=\alpha \mu_{lin}$, wherein $\alpha > 1$ an adjustable coefficient. The amplified sub-threshold mobility is defined in such a way to accommodate the large subthreshold current observed in our simulation. The slope factor $n_{slope}$ is used to modify the subthreshold slope for short-channel devices~\cite{Tang:09}.

\begin{figure}[ht]
\begin{center}
\includegraphics[width=\columnwidth]{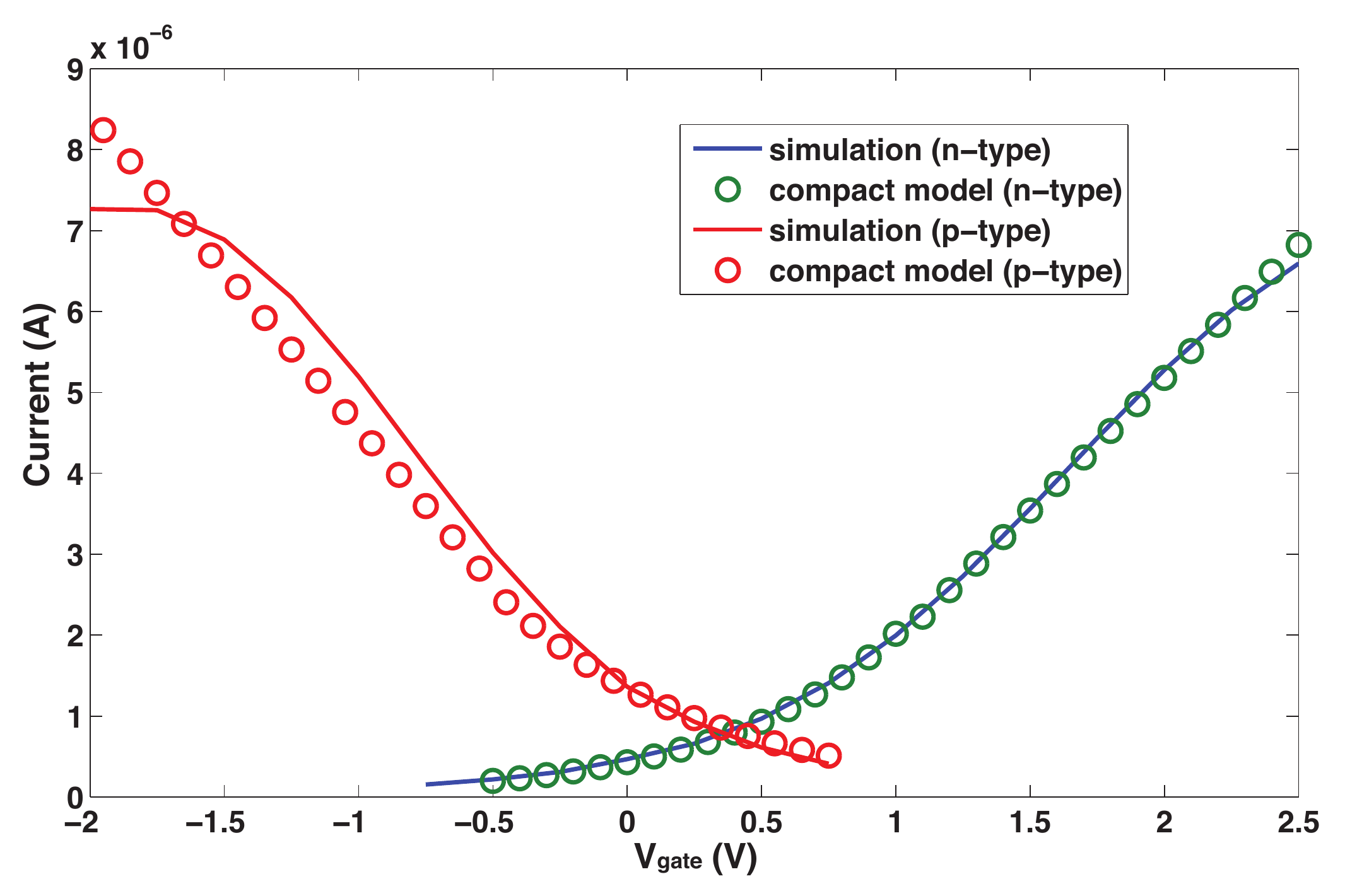}
\caption{ $I-V_{gate}$ curves from QM/EM simulation and compact modeling.\label{fig:IV_compact}}
\end{center}
\end{figure}

\begin{figure}[ht]
\begin{center}
\includegraphics[width=\columnwidth]{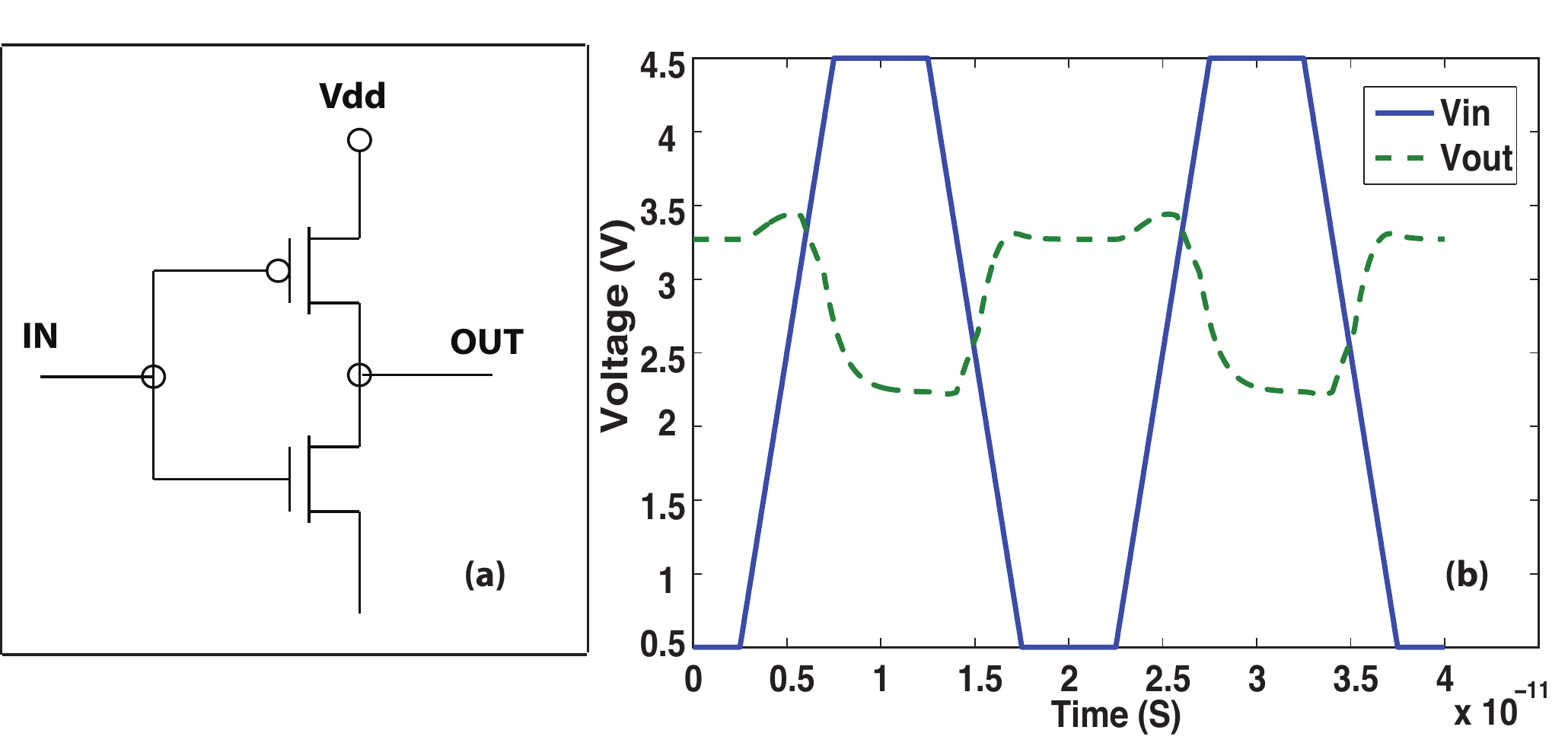}
\caption{(a) Circuit diagram of inverter. (b) Simulated I/O voltages with the generated compact model. } \label{fig:sim_inverter}
\end{center}
\end{figure}

The $I-V_{gate}$ curves in Fig. \ref{fig:current} (a) are fitted into the compact model by using equations (\ref{eq:curr_lin}), (\ref{eq:curr_sat}) and (\ref{eq:curr_sub}), as shown in Fig.~\ref{fig:IV_compact}. In addition, we incorporate the generated compact model into a Matlab-based SPICE-like simulator SMORES~\cite{SMORES} to simulate an inverter circuit. The invertor, as shown in Fig.~\ref{fig:sim_inverter} (a), consists of an n-type and a p-type junctionless transistor. The transient circuit simulation result is shown in Fig.~\ref{fig:sim_inverter} (b). The output waveform is generally ``inverted'' from the input waveform, which verifies the functionality of the inverter. The peak-valley voltage difference in the output is relatively small, due to the low ON/OFF current ratio. It is expected that a higher ON/OFF ratio improves the voltage difference.

\section{Multi-Scale EDA Scheme}
\begin{figure}[ht]
\begin{center}
\includegraphics[width=\columnwidth]{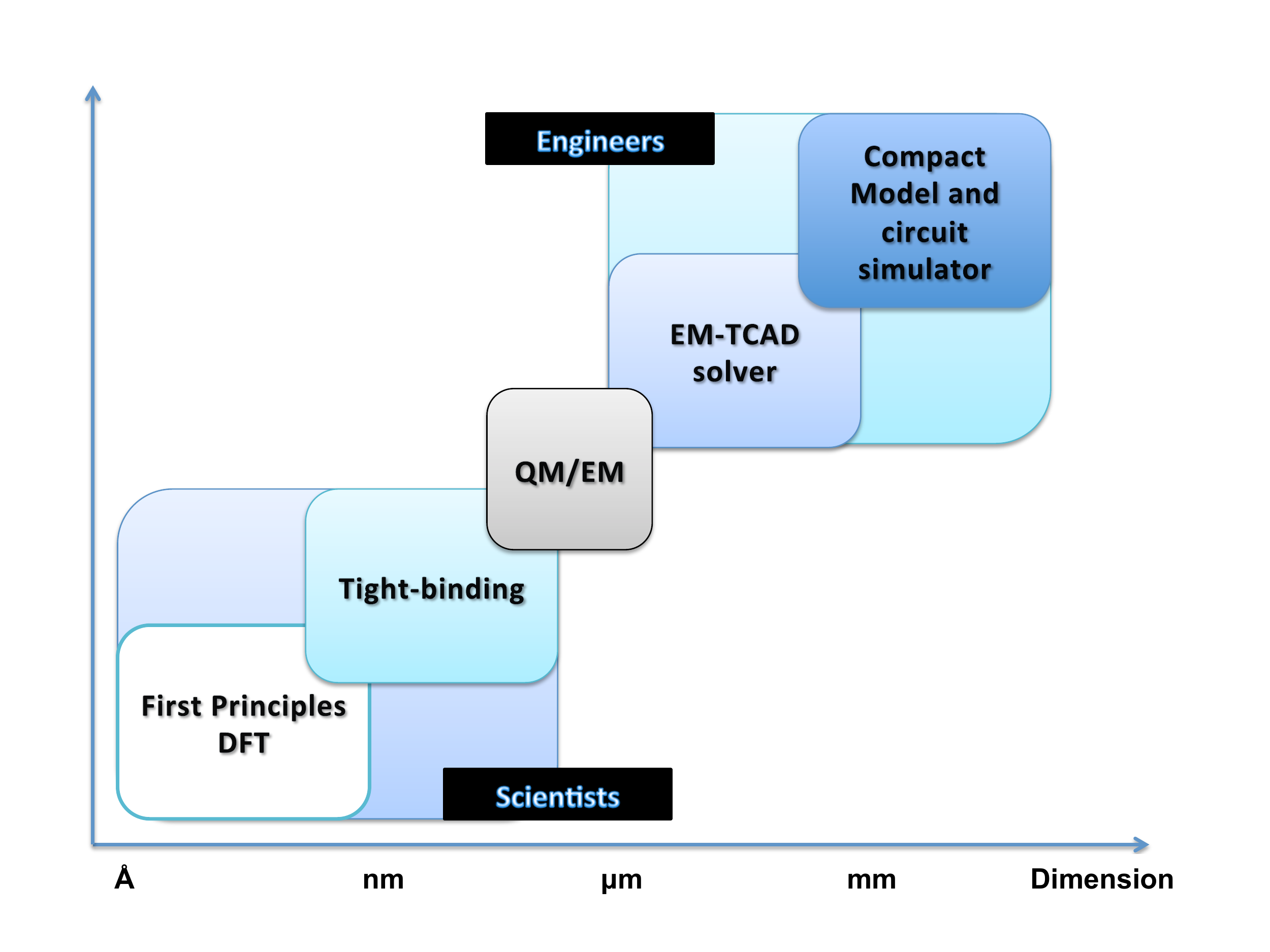}
\caption{Hierarchy of the multi-scale EDA flow.\label{fig:multiscale}}
\end{center}
\end{figure}

Given the creative feature of the research field and the far greater quantum complexity of the nanoscale systems, it becomes very important for computational scientists to work closely with the engineers on developing the next generation multi-scale EDA technology which is proficient in both the nano-scale insights and the application scale. In addition, experiments have become increasingly difficult and expensive, and thus may not be a viable solution in the long term. Recognizing these facts, we devise a multi-scale simulation-based EDA flow as illustrated in Fig.~\ref{fig:multiscale}. The flow starts from the most fundamental ab-initio characterization of nano-scale structures, goes all the way up to the classical modeling of semiconductor devices and compact modeling generation for large-scale SPICE simulation. The QM/EM scheme bridges the two regimes which have long been isolated. The success of this flow enables the first-principles accuracy to come into the chip-level simulation and the engineers' world. In the longer term, the seamless connection between the atomistic first-principles quantum mechanical calculation and the circuit simulation that is bridged by the QM/EM method can be further enhanced by the rapid development of computational power.

\section{Conclusion}

We have presented the development of a new generation of simulation tools with the aim of providing the atomistic level of understanding capability for circuit design and with the hope that these tools may help us to design new devices and circuits. The aim is to package these tools in such a way that any experimentalists, device and circuit design engineers will be able to access the technology without requiring any specialist. Application to the simulation of junctionless transistor has demonstrated in principle the viability of the new EDA flow. The EDA tool starting from atomistic quantum mechanical simulation all the way to integrated circuit design of sub-16nm may thus be possible. Designing electronic devices on computer screen may not be a distant reality.

\begin{acknowledgments}
Support from the Hong Kong Research Grant Council (Contract Nos. HKU7009/09P, 7008/08P, 7007/11P, and HKUST 9/CRF/08), and the University Grant Council (Contract No. AoE/P-04/08) is gratefully acknowledged.
\end{acknowledgments}

\appendix


\begin{thebibliography}{40}%
\makeatletter
\providecommand \@ifxundefined [1]{%
 \@ifx{#1\undefined}
}%
\providecommand \@ifnum [1]{%
 \ifnum #1\expandafter \@firstoftwo
 \else \expandafter \@secondoftwo
 \fi
}%
\providecommand \@ifx [1]{%
 \ifx #1\expandafter \@firstoftwo
 \else \expandafter \@secondoftwo
 \fi
}%
\providecommand \natexlab [1]{#1}%
\providecommand \enquote  [1]{``#1''}%
\providecommand \bibnamefont  [1]{#1}%
\providecommand \bibfnamefont [1]{#1}%
\providecommand \citenamefont [1]{#1}%
\providecommand \href@noop [0]{\@secondoftwo}%
\providecommand \href [0]{\begingroup \@sanitize@url \@href}%
\providecommand \@href[1]{\@@startlink{#1}\@@href}%
\providecommand \@@href[1]{\endgroup#1\@@endlink}%
\providecommand \@sanitize@url [0]{\catcode `\\12\catcode `\$12\catcode
  `\&12\catcode `\#12\catcode `\^12\catcode `\_12\catcode `\%12\relax}%
\providecommand \@@startlink[1]{}%
\providecommand \@@endlink[0]{}%
\providecommand \url  [0]{\begingroup\@sanitize@url \@url }%
\providecommand \@url [1]{\endgroup\@href {#1}{\urlprefix }}%
\providecommand \urlprefix  [0]{URL }%
\providecommand \Eprint [0]{\href }%
\providecommand \doibase [0]{http://dx.doi.org/}%
\providecommand \selectlanguage [0]{\@gobble}%
\providecommand \bibinfo  [0]{\@secondoftwo}%
\providecommand \bibfield  [0]{\@secondoftwo}%
\providecommand \translation [1]{[#1]}%
\providecommand \BibitemOpen [0]{}%
\providecommand \bibitemStop [0]{}%
\providecommand \bibitemNoStop [0]{.\EOS\space}%
\providecommand \EOS [0]{\spacefactor3000\relax}%
\providecommand \BibitemShut  [1]{\csname bibitem#1\endcsname}%
\let\auto@bib@innerbib\@empty
\bibitem [{ITR(2011)}]{ITRS:11}%
  \BibitemOpen
  \href {http://www.itrs.net/Links/2011ITRS/2011Chapters/2011ExecSum.pdf}
  {\enquote {\bibinfo {title} {International technology roadmap for
  semiconductors: Executive summary},}\ }\bibinfo {type} {Tech. Rep.}\
  (\bibinfo  {institution} {International Technology Roadmap for
  Semiconductors},\ \bibinfo {year} {2011})\BibitemShut {NoStop}%
\bibitem [{\citenamefont {Devoret}\ and\ \citenamefont
  {Schoelkopf}(2000)}]{Devoret:2000fk}%
  \BibitemOpen
  \bibfield  {author} {\bibinfo {author} {\bibfnamefont {M.}~\bibnamefont
  {Devoret}}\ and\ \bibinfo {author} {\bibfnamefont {R.}~\bibnamefont
  {Schoelkopf}},\ }\bibfield  {title} {\enquote {\bibinfo {title} {Amplifying
  quantum signals with the single-electron transistor},}\ }\href {\doibase
  10.1038/35023253} {\bibfield  {journal} {\bibinfo  {journal} {Nature}\
  }\textbf {\bibinfo {volume} {406}},\ \bibinfo {pages} {1039--1046} (\bibinfo
  {year} {2000})}\BibitemShut {NoStop}%
\bibitem [{\citenamefont {Cui}\ \emph {et~al.}(2003)\citenamefont {Cui},
  \citenamefont {Zhong}, \citenamefont {Wang}, \citenamefont {Wang},\ and\
  \citenamefont {Lieber}}]{Cui:03}%
  \BibitemOpen
  \bibfield  {author} {\bibinfo {author} {\bibfnamefont {Y.}~\bibnamefont
  {Cui}}, \bibinfo {author} {\bibfnamefont {Z.~h.}\ \bibnamefont {Zhong}},
  \bibinfo {author} {\bibfnamefont {D.}~\bibnamefont {Wang}}, \bibinfo {author}
  {\bibfnamefont {W.~U.}\ \bibnamefont {Wang}}, \ and\ \bibinfo {author}
  {\bibfnamefont {C.~M.}\ \bibnamefont {Lieber}},\ }\bibfield  {title}
  {\enquote {\bibinfo {title} {High performance silicon nanowire field effect
  transistors},}\ }\href@noop {} {\bibfield  {journal} {\bibinfo  {journal}
  {Nano Letters}\ }\textbf {\bibinfo {volume} {3}},\ \bibinfo {pages}
  {149--152} (\bibinfo {year} {2003})}\BibitemShut {NoStop}%
\bibitem [{\citenamefont {Yam}\ \emph {et~al.}(2011)\citenamefont {Yam},
  \citenamefont {Meng}, \citenamefont {Chen}, \citenamefont {Chen},\ and\
  \citenamefont {Wong}}]{yam2011}%
  \BibitemOpen
  \bibfield  {author} {\bibinfo {author} {\bibfnamefont {C.-Y.}\ \bibnamefont
  {Yam}}, \bibinfo {author} {\bibfnamefont {L.-Y.}\ \bibnamefont {Meng}},
  \bibinfo {author} {\bibfnamefont {G.-H.}\ \bibnamefont {Chen}}, \bibinfo
  {author} {\bibfnamefont {Q.}~\bibnamefont {Chen}}, \ and\ \bibinfo {author}
  {\bibfnamefont {N.}~\bibnamefont {Wong}},\ }\bibfield  {title} {\enquote
  {\bibinfo {title} {Multiscale quantum mechanics/electromagnetics simulation
  for electronic device},}\ }\href@noop {} {\bibfield  {journal} {\bibinfo
  {journal} {Phys. Chem. Chem. Phys.}\ }\textbf {\bibinfo {volume} {13}},\
  \bibinfo {pages} {14365--14369} (\bibinfo {year} {2011})}\BibitemShut
  {NoStop}%
\bibitem [{\citenamefont {Meng}\ \emph {et~al.}(2012)\citenamefont {Meng},
  \citenamefont {Yam}, \citenamefont {Koo}, \citenamefont {Chen}, \citenamefont
  {Wong},\ and\ \citenamefont {Chen}}]{meng2012}%
  \BibitemOpen
  \bibfield  {author} {\bibinfo {author} {\bibfnamefont {L.-Y.}\ \bibnamefont
  {Meng}}, \bibinfo {author} {\bibfnamefont {C.-Y.}\ \bibnamefont {Yam}},
  \bibinfo {author} {\bibfnamefont {S.-K.}\ \bibnamefont {Koo}}, \bibinfo
  {author} {\bibfnamefont {Q.}~\bibnamefont {Chen}}, \bibinfo {author}
  {\bibfnamefont {N.}~\bibnamefont {Wong}}, \ and\ \bibinfo {author}
  {\bibfnamefont {G.-H.}\ \bibnamefont {Chen}},\ }\bibfield  {title} {\enquote
  {\bibinfo {title} {Dynamic multiscale quantum mechanics/electromagnetics
  simulation method},}\ }\href@noop {} {\bibfield  {journal} {\bibinfo
  {journal} {J. Chem. Theory Comput.}\ }\textbf {\bibinfo {volume} {8}},\
  \bibinfo {pages} {1190} (\bibinfo {year} {2012})}\BibitemShut {NoStop}%
\bibitem [{\citenamefont {Nagel}(1975)}]{SPICE}%
  \BibitemOpen
  \bibfield  {author} {\bibinfo {author} {\bibfnamefont {L.}~\bibnamefont
  {Nagel}},\ }\href@noop {} {\emph {\bibinfo {title} {{SPICE2: A computer
  program to simulate semiconductor circuits}}}}\ (\bibinfo  {publisher}
  {MIT},\ \bibinfo {year} {1975})\BibitemShut {NoStop}%
\bibitem [{\citenamefont {Chen}(2005)}]{lodestar}%
  \BibitemOpen
  \bibfield  {author} {\bibinfo {author} {\bibfnamefont {G.-H.}\ \bibnamefont
  {Chen}},\ }\href@noop {} {\emph {\bibinfo {title} {{LODESTAR}}}} (\bibinfo
  {year} {2005}),\ \bibinfo {note}
  {\url{http://yangtze.hku.hk/LODESTAR/lodestar.php}}\BibitemShut {NoStop}%
\bibitem [{\citenamefont {Pecchia}\ \emph {et~al.}(2008)\citenamefont
  {Pecchia}, \citenamefont {Penazzi}, \citenamefont {Salvucci},\ and\
  \citenamefont {Di~Carlo}}]{Pecchia:2008fk}%
  \BibitemOpen
  \bibfield  {author} {\bibinfo {author} {\bibfnamefont {A.}~\bibnamefont
  {Pecchia}}, \bibinfo {author} {\bibfnamefont {G.}~\bibnamefont {Penazzi}},
  \bibinfo {author} {\bibfnamefont {L.}~\bibnamefont {Salvucci}}, \ and\
  \bibinfo {author} {\bibfnamefont {A.}~\bibnamefont {Di~Carlo}},\ }\bibfield
  {title} {\enquote {\bibinfo {title} {Non-equilibrium green's functions in
  density functional tight binding: method and applications},}\ }\href
  {\doibase 10.1088/1367-2630/10/6/065022} {\bibfield  {journal} {\bibinfo
  {journal} {New Journal of Physics}\ }\textbf {\bibinfo {volume} {10}},\
  \bibinfo {pages} {065022} (\bibinfo {year} {2008})}\BibitemShut {NoStop}%
\bibitem [{\citenamefont {Meuris}, \citenamefont {Schoenmaker},\ and\
  \citenamefont {Magnus}(2001)}]{Wim:01}%
  \BibitemOpen
  \bibfield  {author} {\bibinfo {author} {\bibfnamefont {P.}~\bibnamefont
  {Meuris}}, \bibinfo {author} {\bibfnamefont {W.}~\bibnamefont {Schoenmaker}},
  \ and\ \bibinfo {author} {\bibfnamefont {W.}~\bibnamefont {Magnus}},\
  }\bibfield  {title} {\enquote {\bibinfo {title} {Strategy for electromagnetic
  interconnect modeling},}\ }\href@noop {} {\bibfield  {journal} {\bibinfo
  {journal} {IEEE Trans. on CAD}\ }\textbf {\bibinfo {volume} {20}},\ \bibinfo
  {pages} {753--762} (\bibinfo {year} {2001})}\BibitemShut {NoStop}%
\bibitem [{\citenamefont {Schoenmaker}\ and\ \citenamefont
  {Meuris}(2002)}]{Wim:02}%
  \BibitemOpen
  \bibfield  {author} {\bibinfo {author} {\bibfnamefont {W.}~\bibnamefont
  {Schoenmaker}}\ and\ \bibinfo {author} {\bibfnamefont {P.}~\bibnamefont
  {Meuris}},\ }\bibfield  {title} {\enquote {\bibinfo {title} {Electromagnetic
  interconnects and passives modeling: software implementation issues},}\
  }\href@noop {} {\bibfield  {journal} {\bibinfo  {journal} {IEEE Trans. on
  CAD}\ }\textbf {\bibinfo {volume} {21}},\ \bibinfo {pages} {534--543}
  (\bibinfo {year} {2002})}\BibitemShut {NoStop}%
\bibitem [{\citenamefont {Lilienfeld}(1925)}]{lilienfeld1925}%
  \BibitemOpen
  \bibfield  {author} {\bibinfo {author} {\bibfnamefont {J.~E.}\ \bibnamefont
  {Lilienfeld}},\ }\bibfield  {title} {\enquote {\bibinfo {title} {Method and
  apparatus for controlling electric current},}\ }\href@noop {} {\bibfield
  {journal} {\bibinfo  {journal} {US patent}\ }\textbf {\bibinfo {volume}
  {1}},\ \bibinfo {pages} {175} (\bibinfo {year} {1925})}\BibitemShut {NoStop}%
\bibitem [{\citenamefont {Sor{\'e}e}, \citenamefont {Magnus},\ and\
  \citenamefont {Pourtois}(2008)}]{soree2008}%
  \BibitemOpen
  \bibfield  {author} {\bibinfo {author} {\bibfnamefont {B.}~\bibnamefont
  {Sor{\'e}e}}, \bibinfo {author} {\bibfnamefont {W.}~\bibnamefont {Magnus}}, \
  and\ \bibinfo {author} {\bibfnamefont {G.}~\bibnamefont {Pourtois}},\
  }\bibfield  {title} {\enquote {\bibinfo {title} {Analytical and
  self-consistent quantum mechanical model for a surrounding gate mos nanowire
  operated in jfet mode},}\ }\href {\doibase 10.1007/s10825-008-0217-3}
  {\bibfield  {journal} {\bibinfo  {journal} {Journal of Computational
  Electronics}\ }\textbf {\bibinfo {volume} {7}},\ \bibinfo {pages} {380--383}
  (\bibinfo {year} {2008})}\BibitemShut {NoStop}%
\bibitem [{\citenamefont {Colinge}\ \emph {et~al.}(2010)\citenamefont
  {Colinge}, \citenamefont {Lee}, \citenamefont {Afzalian}, \citenamefont
  {Akhavan}, \citenamefont {Yan}, \citenamefont {Ferain}, \citenamefont
  {Razavi}, \citenamefont {O'Neill}, \citenamefont {Blake}, \citenamefont
  {White}, \citenamefont {Kelleher}, \citenamefont {McCarthy},\ and\
  \citenamefont {Murphy}}]{colinge2010}%
  \BibitemOpen
  \bibfield  {author} {\bibinfo {author} {\bibfnamefont {J.-P.}\ \bibnamefont
  {Colinge}}, \bibinfo {author} {\bibfnamefont {C.-W.}\ \bibnamefont {Lee}},
  \bibinfo {author} {\bibfnamefont {A.}~\bibnamefont {Afzalian}}, \bibinfo
  {author} {\bibfnamefont {N.}~\bibnamefont {Akhavan}}, \bibinfo {author}
  {\bibfnamefont {R.}~\bibnamefont {Yan}}, \bibinfo {author} {\bibfnamefont
  {I.}~\bibnamefont {Ferain}}, \bibinfo {author} {\bibfnamefont
  {P.}~\bibnamefont {Razavi}}, \bibinfo {author} {\bibfnamefont
  {B.}~\bibnamefont {O'Neill}}, \bibinfo {author} {\bibfnamefont
  {A.}~\bibnamefont {Blake}}, \bibinfo {author} {\bibfnamefont
  {M.}~\bibnamefont {White}}, \bibinfo {author} {\bibfnamefont {A.-M.}\
  \bibnamefont {Kelleher}}, \bibinfo {author} {\bibfnamefont {B.}~\bibnamefont
  {McCarthy}}, \ and\ \bibinfo {author} {\bibfnamefont {R.}~\bibnamefont
  {Murphy}},\ }\bibfield  {title} {\enquote {\bibinfo {title} {Nanowire
  transistors without junctions},}\ }\href {\doibase 10.1038/nnano.2010.15}
  {\bibfield  {journal} {\bibinfo  {journal} {Nature Nanotechnology}\ }\textbf
  {\bibinfo {volume} {5}},\ \bibinfo {pages} {225--229} (\bibinfo {year}
  {2010})}\BibitemShut {NoStop}%
\bibitem [{\citenamefont {Lee}\ \emph {et~al.}(2010{\natexlab{a}})\citenamefont
  {Lee}, \citenamefont {Nazarov}, \citenamefont {Ferain}, \citenamefont
  {Akhavan}, \citenamefont {Yan}, \citenamefont {Razavi}, \citenamefont {Yu},
  \citenamefont {Doria},\ and\ \citenamefont {Colinge}}]{lee2010a}%
  \BibitemOpen
  \bibfield  {author} {\bibinfo {author} {\bibfnamefont {C.}~\bibnamefont
  {Lee}}, \bibinfo {author} {\bibfnamefont {A.}~\bibnamefont {Nazarov}},
  \bibinfo {author} {\bibfnamefont {I.}~\bibnamefont {Ferain}}, \bibinfo
  {author} {\bibfnamefont {N.}~\bibnamefont {Akhavan}}, \bibinfo {author}
  {\bibfnamefont {R.}~\bibnamefont {Yan}}, \bibinfo {author} {\bibfnamefont
  {P.}~\bibnamefont {Razavi}}, \bibinfo {author} {\bibfnamefont
  {R.}~\bibnamefont {Yu}}, \bibinfo {author} {\bibfnamefont {R.}~\bibnamefont
  {Doria}}, \ and\ \bibinfo {author} {\bibfnamefont {J.}~\bibnamefont
  {Colinge}},\ }\bibfield  {title} {\enquote {\bibinfo {title} {Low
  subthreshold slope in junctionless multigate transistors},}\ }\href {\doibase
  10.1063/1.3358131} {\bibfield  {journal} {\bibinfo  {journal} {Applied
  Physics Letters}\ }\textbf {\bibinfo {volume} {96}},\ \bibinfo {pages}
  {102106} (\bibinfo {year} {2010}{\natexlab{a}})}\BibitemShut {NoStop}%
\bibitem [{\citenamefont {Lee}\ \emph {et~al.}(2010{\natexlab{b}})\citenamefont
  {Lee}, \citenamefont {Ferain}, \citenamefont {Afzalian}, \citenamefont {Yan},
  \citenamefont {Akhavan}, \citenamefont {Razavi},\ and\ \citenamefont
  {Colinge}}]{lee2010b}%
  \BibitemOpen
  \bibfield  {author} {\bibinfo {author} {\bibfnamefont {C.-W.}\ \bibnamefont
  {Lee}}, \bibinfo {author} {\bibfnamefont {I.}~\bibnamefont {Ferain}},
  \bibinfo {author} {\bibfnamefont {A.}~\bibnamefont {Afzalian}}, \bibinfo
  {author} {\bibfnamefont {R.}~\bibnamefont {Yan}}, \bibinfo {author}
  {\bibfnamefont {N.}~\bibnamefont {Akhavan}}, \bibinfo {author} {\bibfnamefont
  {P.}~\bibnamefont {Razavi}}, \ and\ \bibinfo {author} {\bibfnamefont {J.-P.}\
  \bibnamefont {Colinge}},\ }\bibfield  {title} {\enquote {\bibinfo {title}
  {Performance estimation of junctionless multigate transistors},}\ }\href
  {\doibase 10.1016/j.sse.2009.12.003} {\bibfield  {journal} {\bibinfo
  {journal} {Solid-State Electronics}\ }\textbf {\bibinfo {volume} {54}},\
  \bibinfo {pages} {97--103} (\bibinfo {year}
  {2010}{\natexlab{b}})}\BibitemShut {NoStop}%
\bibitem [{\citenamefont {Raskin}\ \emph {et~al.}(2010)\citenamefont {Raskin},
  \citenamefont {Colinge}, \citenamefont {Ferain}, \citenamefont {Kranti},
  \citenamefont {Lee}, \citenamefont {Akhavan}, \citenamefont {Yan},
  \citenamefont {Razavi},\ and\ \citenamefont {Yu}}]{Raskin}%
  \BibitemOpen
  \bibfield  {author} {\bibinfo {author} {\bibfnamefont {J.-P.}\ \bibnamefont
  {Raskin}}, \bibinfo {author} {\bibfnamefont {J.-P.}\ \bibnamefont {Colinge}},
  \bibinfo {author} {\bibfnamefont {I.}~\bibnamefont {Ferain}}, \bibinfo
  {author} {\bibfnamefont {A.}~\bibnamefont {Kranti}}, \bibinfo {author}
  {\bibfnamefont {C.-W.}\ \bibnamefont {Lee}}, \bibinfo {author} {\bibfnamefont
  {N.~D.}\ \bibnamefont {Akhavan}}, \bibinfo {author} {\bibfnamefont
  {R.}~\bibnamefont {Yan}}, \bibinfo {author} {\bibfnamefont {P.}~\bibnamefont
  {Razavi}}, \ and\ \bibinfo {author} {\bibfnamefont {R.}~\bibnamefont {Yu}},\
  }\bibfield  {title} {\enquote {\bibinfo {title} {Mobility improvement in
  nanowire junctionless transistors by uniaxial strain},}\ }\href@noop {}
  {\bibfield  {journal} {\bibinfo  {journal} {APPLIED PHYSICS LETTERS}\
  }\textbf {\bibinfo {volume} {97}},\ \bibinfo {pages} {042114} (\bibinfo
  {year} {2010})}\BibitemShut {NoStop}%
\bibitem [{\citenamefont {Akhavan}\ \emph {et~al.}(2011)\citenamefont
  {Akhavan}, \citenamefont {Ferain}, \citenamefont {Razavi}, \citenamefont
  {Yu},\ and\ \citenamefont {Colinge}}]{Akhavan2011}%
  \BibitemOpen
  \bibfield  {author} {\bibinfo {author} {\bibfnamefont {N.~D.}\ \bibnamefont
  {Akhavan}}, \bibinfo {author} {\bibfnamefont {I.}~\bibnamefont {Ferain}},
  \bibinfo {author} {\bibfnamefont {P.}~\bibnamefont {Razavi}}, \bibinfo
  {author} {\bibfnamefont {R.}~\bibnamefont {Yu}}, \ and\ \bibinfo {author}
  {\bibfnamefont {J.-P.}\ \bibnamefont {Colinge}},\ }\bibfield  {title}
  {\enquote {\bibinfo {title} {Improvement of carrier ballisticity in
  junctionless nanowire transistors},}\ }\href {\doibase 10.1063/1.3559625}
  {\bibfield  {journal} {\bibinfo  {journal} {Applied Physics Letters}\
  }\textbf {\bibinfo {volume} {98}},\ \bibinfo {pages} {103510--103510--3}
  (\bibinfo {year} {2011})}\BibitemShut {NoStop}%
\bibitem [{\citenamefont {Yan}\ \emph {et~al.}(2011)\citenamefont {Yan},
  \citenamefont {Kranti}, \citenamefont {Ferain}, \citenamefont {Lee},
  \citenamefont {Yu}, \citenamefont {Dehdashti},\ and\ \citenamefont
  {Colinge}}]{Yan2011}%
  \BibitemOpen
  \bibfield  {author} {\bibinfo {author} {\bibfnamefont {R.}~\bibnamefont
  {Yan}}, \bibinfo {author} {\bibfnamefont {A.}~\bibnamefont {Kranti}},
  \bibinfo {author} {\bibfnamefont {I.}~\bibnamefont {Ferain}}, \bibinfo
  {author} {\bibfnamefont {C.-W.}\ \bibnamefont {Lee}}, \bibinfo {author}
  {\bibfnamefont {R.}~\bibnamefont {Yu}}, \bibinfo {author} {\bibfnamefont
  {N.}~\bibnamefont {Dehdashti}}, \ and\ \bibinfo {author} {\bibfnamefont
  {P.}~\bibnamefont {Colinge}},\ }\bibfield  {title} {\enquote {\bibinfo
  {title} {Investigation of high-performance sub-50nm junctionless nanowire
  transistors},}\ }\href {\doibase 10.1016/j.microrel.2011.02.016} {\bibfield
  {journal} {\bibinfo  {journal} {Microelectronics Reliability}\ }\textbf
  {\bibinfo {volume} {51}},\ \bibinfo {pages} {1166--1171} (\bibinfo {year}
  {2011})}\BibitemShut {NoStop}%
\bibitem [{\citenamefont {Ansari}\ \emph {et~al.}(2010)\citenamefont {Ansari},
  \citenamefont {Feldman}, \citenamefont {Fagas}, \citenamefont {Colinge},\
  and\ \citenamefont {Greer}}]{Ansari2010}%
  \BibitemOpen
  \bibfield  {author} {\bibinfo {author} {\bibfnamefont {L.}~\bibnamefont
  {Ansari}}, \bibinfo {author} {\bibfnamefont {B.}~\bibnamefont {Feldman}},
  \bibinfo {author} {\bibfnamefont {G.}~\bibnamefont {Fagas}}, \bibinfo
  {author} {\bibfnamefont {J.-P.}\ \bibnamefont {Colinge}}, \ and\ \bibinfo
  {author} {\bibfnamefont {J.~C.}\ \bibnamefont {Greer}},\ }\bibfield  {title}
  {\enquote {\bibinfo {title} {Simulation of junctionless si nanowire
  transistors with 3 nm gate length},}\ }\href {\doibase 10.1063/1.3478012}
  {\bibfield  {journal} {\bibinfo  {journal} {Applied Physics Letters}\
  }\textbf {\bibinfo {volume} {97}},\ \bibinfo {pages} {062105--062107}
  (\bibinfo {year} {2010})}\BibitemShut {NoStop}%
\bibitem [{\citenamefont {Ansari}\ \emph {et~al.}(2011)\citenamefont {Ansari},
  \citenamefont {Feldman}, \citenamefont {Fagas}, \citenamefont {Colinge},\
  and\ \citenamefont {Greer}}]{Ansari2011}%
  \BibitemOpen
  \bibfield  {author} {\bibinfo {author} {\bibfnamefont {L.}~\bibnamefont
  {Ansari}}, \bibinfo {author} {\bibfnamefont {B.}~\bibnamefont {Feldman}},
  \bibinfo {author} {\bibfnamefont {G.}~\bibnamefont {Fagas}}, \bibinfo
  {author} {\bibfnamefont {J.-P.}\ \bibnamefont {Colinge}}, \ and\ \bibinfo
  {author} {\bibfnamefont {J.}~\bibnamefont {Greer}},\ }\bibfield  {title}
  {\enquote {\bibinfo {title} {Subthreshold behavior of junctionless silicon
  nanowire transistors from atomic scale simulations},}\ }\href@noop {}
  {\bibfield  {journal} {\bibinfo  {journal} {Solid-State Electronics}\
  }\textbf {\bibinfo {volume} {71}},\ \bibinfo {pages} {58} (\bibinfo {year}
  {2011})}\BibitemShut {NoStop}%
\bibitem [{\citenamefont {Martinez}\ \emph {et~al.}(2011)\citenamefont
  {Martinez}, \citenamefont {Aldegunde}, \citenamefont {Brown}, \citenamefont
  {Roy},\ and\ \citenamefont {Asenov}}]{Martinez2011}%
  \BibitemOpen
  \bibfield  {author} {\bibinfo {author} {\bibfnamefont {A.}~\bibnamefont
  {Martinez}}, \bibinfo {author} {\bibfnamefont {M.}~\bibnamefont {Aldegunde}},
  \bibinfo {author} {\bibfnamefont {A.}~\bibnamefont {Brown}}, \bibinfo
  {author} {\bibfnamefont {S.}~\bibnamefont {Roy}}, \ and\ \bibinfo {author}
  {\bibfnamefont {A.}~\bibnamefont {Asenov}},\ }\bibfield  {title} {\enquote
  {\bibinfo {title} {Negf simulations of a junctionless si gate-all-around
  nanowire transistor with discrete dopants},}\ }\href {\doibase
  10.1016/j.sse.2011.10.028} {\bibfield  {journal} {\bibinfo  {journal}
  {Solid-State Electronics}\ ,\ \bibinfo {pages} {101--105}} (\bibinfo {year}
  {2011})}\BibitemShut {NoStop}%
\bibitem [{\citenamefont {Gnani}\ \emph {et~al.}(2011)\citenamefont {Gnani},
  \citenamefont {Gnudi}, \citenamefont {Reggiani}, \citenamefont {Baccarani},
  \citenamefont {Shen}, \citenamefont {Singh}, \citenamefont {Lo},\ and\
  \citenamefont {Kwong}}]{Gnani2011}%
  \BibitemOpen
  \bibfield  {author} {\bibinfo {author} {\bibfnamefont {E.}~\bibnamefont
  {Gnani}}, \bibinfo {author} {\bibfnamefont {A.}~\bibnamefont {Gnudi}},
  \bibinfo {author} {\bibfnamefont {S.}~\bibnamefont {Reggiani}}, \bibinfo
  {author} {\bibfnamefont {G.}~\bibnamefont {Baccarani}}, \bibinfo {author}
  {\bibfnamefont {N.}~\bibnamefont {Shen}}, \bibinfo {author} {\bibfnamefont
  {N.}~\bibnamefont {Singh}}, \bibinfo {author} {\bibfnamefont
  {G.}~\bibnamefont {Lo}}, \ and\ \bibinfo {author} {\bibfnamefont
  {D.}~\bibnamefont {Kwong}},\ }\bibfield  {title} {\enquote {\bibinfo {title}
  {Numerical investigation on the junctionless nanowire fet},}\ }\href
  {\doibase 10.1016/j.sse.2011.10.013} {\bibfield  {journal} {\bibinfo
  {journal} {Solid-State Electronics}\ ,\ \bibinfo {pages} {13--18}} (\bibinfo
  {year} {2011})}\BibitemShut {NoStop}%
\bibitem [{\citenamefont {Sels}, \citenamefont {Sor{\'e}e},\ and\ \citenamefont
  {Groeseneken}(2011)}]{Sels2011}%
  \BibitemOpen
  \bibfield  {author} {\bibinfo {author} {\bibfnamefont {D.}~\bibnamefont
  {Sels}}, \bibinfo {author} {\bibfnamefont {B.}~\bibnamefont {Sor{\'e}e}}, \
  and\ \bibinfo {author} {\bibfnamefont {G.}~\bibnamefont {Groeseneken}},\
  }\bibfield  {title} {\enquote {\bibinfo {title} {Quantum ballistic transport
  in the junctionless nanowire pinch-off field effect transistor},}\ }\href
  {\doibase 10.1007/s10825-011-0350-2} {\bibfield  {journal} {\bibinfo
  {journal} {Journal of Computational Electronics}\ ,\ \bibinfo {pages} {1--6}}
  (\bibinfo {year} {2011})}\BibitemShut {NoStop}%
\bibitem [{\citenamefont {Pham}\ \emph {et~al.}(2011)\citenamefont {Pham},
  \citenamefont {Sor{\'e}e}, \citenamefont {Magnus}, \citenamefont {Jungemann},
  \citenamefont {Meinerzhagen},\ and\ \citenamefont {Pourtois}}]{Pham2011}%
  \BibitemOpen
  \bibfield  {author} {\bibinfo {author} {\bibfnamefont {A.-T.}\ \bibnamefont
  {Pham}}, \bibinfo {author} {\bibfnamefont {B.}~\bibnamefont {Sor{\'e}e}},
  \bibinfo {author} {\bibfnamefont {W.}~\bibnamefont {Magnus}}, \bibinfo
  {author} {\bibfnamefont {C.}~\bibnamefont {Jungemann}}, \bibinfo {author}
  {\bibfnamefont {B.}~\bibnamefont {Meinerzhagen}}, \ and\ \bibinfo {author}
  {\bibfnamefont {G.}~\bibnamefont {Pourtois}},\ }\bibfield  {title} {\enquote
  {\bibinfo {title} {Quantum simulations of electrostatics in si cylindrical
  junctionless nanowire nfets and pfets with a homogeneous channel including
  strain and arbitrary crystallographic orientations},}\ }\href {\doibase
  10.1016/j.sse.2011.10.016} {\bibfield  {journal} {\bibinfo  {journal}
  {Solid-State Electronics}\ }\textbf {\bibinfo {volume} {71}},\ \bibinfo
  {pages} {30--36} (\bibinfo {year} {2011})}\BibitemShut {NoStop}%
\bibitem [{\citenamefont {Park}\ \emph {et~al.}(2012)\citenamefont {Park},
  \citenamefont {Ko}, \citenamefont {Kim}, \citenamefont {Baek}, \citenamefont
  {Sohn}, \citenamefont {Baek}, \citenamefont {Park}, \citenamefont {Deen},
  \citenamefont {Jeong},\ and\ \citenamefont {Lee}}]{park:2012qr}%
  \BibitemOpen
  \bibfield  {author} {\bibinfo {author} {\bibfnamefont {C.-H.}\ \bibnamefont
  {Park}}, \bibinfo {author} {\bibfnamefont {M.-D.}\ \bibnamefont {Ko}},
  \bibinfo {author} {\bibfnamefont {K.-H.}\ \bibnamefont {Kim}}, \bibinfo
  {author} {\bibfnamefont {R.-H.}\ \bibnamefont {Baek}}, \bibinfo {author}
  {\bibfnamefont {C.-W.}\ \bibnamefont {Sohn}}, \bibinfo {author}
  {\bibfnamefont {C.}~\bibnamefont {Baek}}, \bibinfo {author} {\bibfnamefont
  {S.}~\bibnamefont {Park}}, \bibinfo {author} {\bibfnamefont {M.}~\bibnamefont
  {Deen}}, \bibinfo {author} {\bibfnamefont {Y.-H.}\ \bibnamefont {Jeong}}, \
  and\ \bibinfo {author} {\bibfnamefont {J.-S.}\ \bibnamefont {Lee}},\
  }\bibfield  {title} {\enquote {\bibinfo {title} {Electrical characteristics
  of 20-nm junctionless si nanowire transistors},}\ }\href {\doibase
  10.1016/j.sse.2011.11.032} {\bibfield  {journal} {\bibinfo  {journal}
  {Solid-State Electronics}\ }\textbf {\bibinfo {volume} {73}},\ \bibinfo
  {pages} {7--10} (\bibinfo {year} {2012})}\BibitemShut {NoStop}%
\bibitem [{\citenamefont {Duarte}, \citenamefont {Choi},\ and\ \citenamefont
  {Choi}(2011)}]{Duarte:11}%
  \BibitemOpen
  \bibfield  {author} {\bibinfo {author} {\bibfnamefont {J.~P.}\ \bibnamefont
  {Duarte}}, \bibinfo {author} {\bibfnamefont {S.-J.}\ \bibnamefont {Choi}}, \
  and\ \bibinfo {author} {\bibfnamefont {Y.-K.}\ \bibnamefont {Choi}},\
  }\bibfield  {title} {\enquote {\bibinfo {title} {A full-range drain current
  model for double-gate junctionless transistors},}\ }\href@noop {} {\bibfield
  {journal} {\bibinfo  {journal} {IEEE Trans. on Electron Devices}\ }\textbf
  {\bibinfo {volume} {58}},\ \bibinfo {pages} {4219 --4225} (\bibinfo {year}
  {2011})}\BibitemShut {NoStop}%
\bibitem [{\citenamefont {Tang}\ \emph {et~al.}(2009)\citenamefont {Tang},
  \citenamefont {Pregaldiny}, \citenamefont {Lallement},\ and\ \citenamefont
  {Sallese}}]{Tang:09}%
  \BibitemOpen
  \bibfield  {author} {\bibinfo {author} {\bibfnamefont {M.}~\bibnamefont
  {Tang}}, \bibinfo {author} {\bibfnamefont {F.}~\bibnamefont {Pregaldiny}},
  \bibinfo {author} {\bibfnamefont {C.}~\bibnamefont {Lallement}}, \ and\
  \bibinfo {author} {\bibfnamefont {J.-M.}\ \bibnamefont {Sallese}},\
  }\bibfield  {title} {\enquote {\bibinfo {title} {Explicit compact model for
  ultranarrow body finfets},}\ }\href@noop {} {\bibfield  {journal} {\bibinfo
  {journal} {IEEE Trans. on Electron Devices}\ }\textbf {\bibinfo {volume}
  {56}},\ \bibinfo {pages} {1543 --1547} (\bibinfo {year} {2009})}\BibitemShut
  {NoStop}%
\bibitem [{\citenamefont {Bond}(2010)}]{SMORES}%
  \BibitemOpen
  \bibfield  {author} {\bibinfo {author} {\bibfnamefont {B.~N.}\ \bibnamefont
  {Bond}},\ }\href@noop {} {\emph {\bibinfo {title} {{SMORES}: A Matlab tool
  for Simulation and Model Order Reduction of Electrical Systems}}} (\bibinfo
  {year} {2010}),\ \bibinfo {note}
  {\url{http://bnbond.com/software/smores/}}\BibitemShut {NoStop}%
\end{thebibliography}
\end{document}